\documentclass[aps,prb,twocolumn,superscriptaddress,showpacs,amsmath,floatfix,citeautoscript]{revtex4-2}

\usepackage[utf8]{inputenc}
\usepackage{graphicx}
\usepackage{mathtools}
\usepackage{xcolor}
\usepackage{amsmath}
\usepackage{amssymb}
\usepackage{xfrac}
\usepackage{soul}

\begin{document}

\title{Tuning of Berry Curvature Dipole in TaAs slabs: An effective Route to Enhance Nonlinear Hall Response}
	
	\author{Hongsheng Pang}
	\affiliation{Key Laboratory of Quantum Information, University of Science and
		Technology of China, Hefei, Anhui, 230026, People's Republic of China}
	
	\author{Gan Jin}
	\affiliation{Key Laboratory of Quantum Information, University of Science and
		Technology of China, Hefei, Anhui, 230026, People's Republic of China}
	
	\author{Lixin He}
	\email[Corresponding author: ]{helx@ustc.edu.cn}
	\affiliation{Key Laboratory of Quantum Information, University of Science and
		Technology of China, Hefei, Anhui, 230026, People's Republic of China}
	\affiliation{Institute of Artificial Intelligence, Hefei Comprehensive National Science Center,
		Hefei, Anhui, 230026, People's Republic of China}
	\affiliation{Hefei National Laboratory, University of Science and
		Technology of China, Hefei, Anhui,  230088, People's Republic of China}
	
\begin{abstract}		
In materials without inversion symmetry, Berry curvature dipole (BCD) arises from the uneven  distribution of Berry curvature in momentum space. This leads to nonlinear anomalous Hall effects even in systems with preserved time-reversal symmetry.
A key goal is to engineer systems with prominent BCD near the Fermi level. Notably, TaAs, a type-I Weyl semimetal, exhibits substantial Berry curvature but a small BCD around the Fermi level.
In this study, we employed first-principles methods to comprehensively investigate the BCD in TaAs. Our findings reveal significant cancellation effects not only within individual Weyl points but crucially, among distinct Weyl point pairs in bulk TaAs. We propose a strategic approach to enhance the BCD in TaAs by employing a layer-stacking technique. This greatly amplifies the BCD compared to the bulk material. By tuning the number of slab layers, we can selectively target specific Weyl point pairs near the Fermi level, while quantum confinement effects suppress contributions from other pairs, mitigating cancellation effects. Especially, the BCD of an 8-layer TaAs slab surpasses the bulk value near the Fermi level by  orders of magnitude.
\end{abstract}	

\date{\today}	
\maketitle
	
\section{Introduction}
	
Since its discovery, the family of Hall effects has garnered significant attention in both scientific research and practical applications.
In contrast to classical understandings of extrinsic magnetization, modern studies have linked Hall effects to band geometry in time-reversal (TR) asymmetric systems \cite{Hall1, Hall2}. However, recent research has shown that even in TR symmetric materials, a nonlinear Hall response can arise when inversion symmetry is broken \cite{BCD_FuLiang,deyo2009semiclassical,PhysRevLett.105.026805} .
This intriguing effect is closely related to topological properties, specifically the Berry curvature dipole (BCD),  which embodies an uneven distribution of Berry curvature within momentum space. Consequently, the BCD can serve as a valuable probe of band geometry in TR systems.
The significance of BCD extends beyond pure academic research.
It possesses effects like frequency doubling and rectification\cite{NonlinearHallRev1, NonlinearHallRev1}, making it highly valuable for advanced electronic devices operating in the gigahertz or terahertz frequency range\cite{THz, piezodevice}.
	
The BCD induced nonlinear Hall effect has been investigated both experimentally and theoretically across diverse material categories,
including transition metal dichalcogenides (TMDCs)~\cite{TMDC,monolayerWTe2,electric_WTe2,bilayerWTe,FewlayerWTe2},  graphene~\cite{twisted-graphene1,twisted-graphene1}, antiferromagnetics~\cite{CuMnAs,AFM}, piezoelectrics\cite{LiOsO,strain-piezo,piezodevice},  among others. Various proposals have been put forth to enhance the BCD in diverse material systems\cite{BCD_Te,LiOsO,BiTeI,TaAs-BCD,FewlayerWTe2,twisted_bilayerWSe2,strainedgraphene,strained-WSe2,bilayerWTe,corrugated_bilayergraphene,twisted_bilayerWTe2}. 	

A promising avenue to enhance BCD involves finding materials with substantial Berry curvature proximate to the Fermi surface, prompting investigations that prioritize topologically nontrivial materials for their distinctive Berry curvature properties\cite{BiTeI}. This pursuit is particularly salient in the context of Weyl semimetals (WSMs) \cite{model}, characterized by pronounced Berry curvature in the vicinity of Weyl points~\cite{TaAs-BCD,THz,TaIrTe4}. 	TaAs, as a representative type-I WSM,  has been examined, revealing the presence of multiple groups of Weyl points proximal to the Fermi level.  Nonetheless, investigations have revealed that the BCD is quite small in the Fermi level vicinity\cite{TaAs-BCD}.
This has been attributed to the mutual cancellation of Berry curvature within individual Weyl points, due to the  type-I nature of the Weyl cone.

In this study, we carry out comprehensive investigation of BCD in TaAs systems using first-principles methods. Our calculations reveals that, apart from the inherent cancelation effects within individual Weyl points,  substantial canceling effects stemming from distinct Weyl point pairs
in bulk TaAs, leading to very small BCD around the Fermi level. We propose a strategic solution for enhancing the BCD in TaAs, employing TaAs slabs, which significantly amplifies the BCD compared to its bulk counterpart.
By tuning the number of stacked layers, we can selectively target specific Weyl point pairs located near the Fermi level, while quantum confinement effects suppress contributions from other pairs, thus mitigating cancelation effects.
Especially, the BCD of an 8-layer TaAs slab surpasses the bulk value around the Fermi level by  orders of magnitude.

\section{Methods}
\label{sec:methods}	
	
We perform first-principle calculations based on density functional theory (DFT) implemented in the Atomic Orbital Based Ab-initio Computation at UStc (ABACUS) package\cite{ABACUS1,ABACUS2}. The ABACUS code specializes in large-scale DFT calculations using the numerical atomic orbital (NAO) basis\cite{orbital,Linpz2023}. The Perdew–Burke–Ernzerhof (PBE) generalized gradient approximation (GGA)\cite{GGA} is employed for the exchange-correlation functional in the calculations.  The SG15\cite{SG15} optimized norm-conserving Vanderbilt pseudopotentials (ONCV)\cite{ONCV} are used, where the Ta $5d^36s^2$ and As $3d^{10}4s^{2}4p^{3}$ electrons are treated as valence electrons. The 2s2p2d1f
NAO bases are used for both Ta and As elements.
In the TaAs slab calculations, a 30 \AA~ vacuum is added to ensure sufficient separation between the upper and lower surfaces in two
adjacent unit cells.
During the self-consistent calculations, a $16\times16\times1$ Monkhorst-Pack $k$-point mesh is used for slabs, whereas a $8\times8\times8$
 $k$-point mesh is used for bulk. The energy cut-off for wave functions is set to 100 Ry. Spin-orbit coupling is taken into account in our calculations.
	
The band structures and BCD are calculated using the PYATB code\cite{pyatb},
which utilizes the tight-binding Hamiltonian directly generated from ABACUS self-consistent calculations.
The BCD can be calculated using the following equation:
	\begin{equation}
		D_{ab}(\mu,T) = \int [d\mathbf{k}] \sum_{n} \frac{\partial E_n}{\partial k_a} \Omega_{n,b}\left(-\frac{\partial f(T,\mu)}{\partial E}\right)_{E=E_n}
		\label{eq:Dab}
	\end{equation}
Here, $f(T,\mu)$ represents the Fermi distribution with a chemical potential $\mu$ at temperature $T$. The crystal directions are denoted by $a$, $b$, and $c$. The Berry curvature $\Omega_{n,b}$ is obtained by the method developed in Ref.\cite{Jin_BerryCurvature}. The smearing constant $k_B T = 2  \text{meV}$ is utilized for the BCD calculations.
The integration $[dk]$ is given by $\frac{d^3 \mathbf{k}}{(2 \pi)^3}$ for a 3D system and $\frac{d^2 \mathbf{k}}{(2 \pi)^2}$ for a 2D system. For the 3D system, a mesh of $500 \times 500 \times 500$ is utilized for integration, and for high contribution points, the grid is further refined with a $20 \times 20 \times 20$ mesh. As for the 2D systems, the integration is carried out on a $500 \times 500 \times 1$ mesh with a $20 \times 20 \times 1$ refinement for important $k$ points.
	
	The expression for the single-band Berry curvature is given by:
	\begin{equation}
		\Omega_{n,c}(\mathbf{k}) = -2 \epsilon_{abc} \text{Im} \sum_{m\neq n} r_{nm,a}(\mathbf{k})r_{mn,b}(\mathbf{k})\, ,
	\end{equation}
where the $r$ matrices represent the connection matrices.
In the context of a 2D system, it is necessary to consider certain aspects. According to the Berry curvature definition \cite{NiuBerry}, the $\frac{\partial}{\partial k_z}$ term is not applicable for a 2D system, leading to $\Omega_x$ and $\Omega_y$ becoming forbidden. Consequently, the Berry curvature becomes a pseudo-scalar with only $\Omega_z$ being present, and $D_{xy}$ should be deemed 0 \cite{BCD_FuLiang}.
	However, in real 2D materials like a TaAs slab, which has finite thickness, the $r_z(\mathbf{k})$ matrix can be interpreted as a Fourier transformation of the real-space connection matrix $r_z(\mathbf{R})$. As a result, $\Omega_x$ and $\Omega_y$ can still exist and be non-zero in this scenario.
	
	In order to quantify the canceling effects present in the systems, we introduce a canceling ratio denoted as $\chi$, which is defined as,
	\begin{equation}
		\chi = \frac{|\int[d\mathbf{k}]D_k|}{\int[d\mathbf{k}]|D_k|} \, ,
		\label{eq:chi}
	\end{equation}
	where, $D_k$ represents the integral of the integrand function in Eq. \ref{eq:Dab} at a given $k$ point.
	The canceling ratio $\chi$ quantifies the degree of canceling effects between contributions from different regions. A value of $\chi$ close to 1 suggests minimal canceling effects, indicating that the contributions largely reinforce each other. On the other hand, a small value of $\chi$ implies significant canceling effects, signifying a substantial reduction in contributions from various regions.

	\section{Results}
	\subsection{TaAs in Bulk State}
	
	\begin{figure}[tbp]
		\centering
		\includegraphics[width=0.4\textwidth]{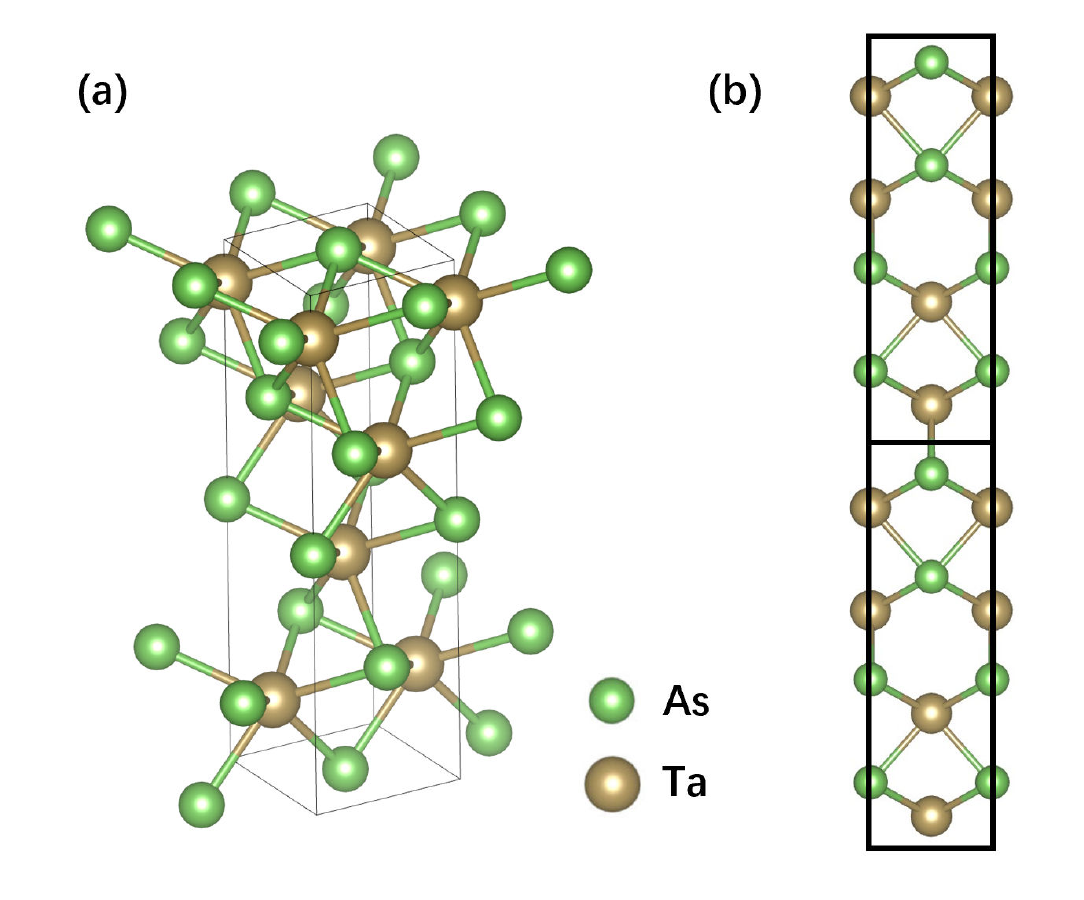}
		\caption{ (a) The crystal structure of bulk TaAs; and (b) an illustration of the structure of an 8-layer TaAs slab.}
		\label{fig:structure}
	\end{figure}
	
	The structure of bulk TaAs is shown in Fig.~\ref{fig:structure}(a), which has the space group $I4_1md$ (No. 109). The structure lacks
	the inversion symmetry but possesses $M_x$ and $M_y$ reflection symmetries \cite{TaAs-WSM, TaAs-experiment}.
	These symmetries play a crucial role in analyzing the properties of the BCD tensor, and guarantee it has  non-zero BCD \cite{TaAs-BCD}.
	
The band structure of bulk TaAs was calculated using density functional theory (DFT), and the results are consistent with previous works \cite{TaAs-WSM, TaAs-experiment, TaAs-fermiarc}. In the absence of spin-orbit coupling (SOC), the band structures have nodal lines which are protected by mirror symmetry. There are two nodal rings in the $k_x = 0$ plane and two in $k_y = 0$ plane. However,  after turning on SOC, the nodal lines break into distinct Weyl points\cite{TaAs-WSM,TaAs-nc}.
The Weyl points are slightly shifted away from the high symmetry lines, and thus the high-symmetry lines are full gapped.
The bulk TaAs system manifests three groups of Weyl points, in which W$_1$ and W$_2$, situate at energy levels 23 meV and 14 meV below the Fermi energy, respectively, and a third group of Weyl points, W$_3$, are 75 meV above the Fermi energy \cite{TaAs-BCD, TaAs-nc}. The C$_{4v}$ symmetry and time reversal symmetry preserve four pairs of W$_1$ points on $k_z = 0$ plane, eight pairs of W$_2$ on $k_z = \pm 0.408$ and eight pairs of W$_3$ on $k_z = \pm0.452$, each pair with two Weyl points of opposite chirality\cite{TaAs-WSM,TaAs-nc} related by the time reversal symmetry near $k_x$, $k_y = 0$ planes. Among them, W$_1$  and W$_2$ are type-I Weyl points and W$_3$ are type-II Weyl points\cite{TaAs-BCD}.

	\begin{figure}[tb]
		\includegraphics[width=0.4 \textwidth]{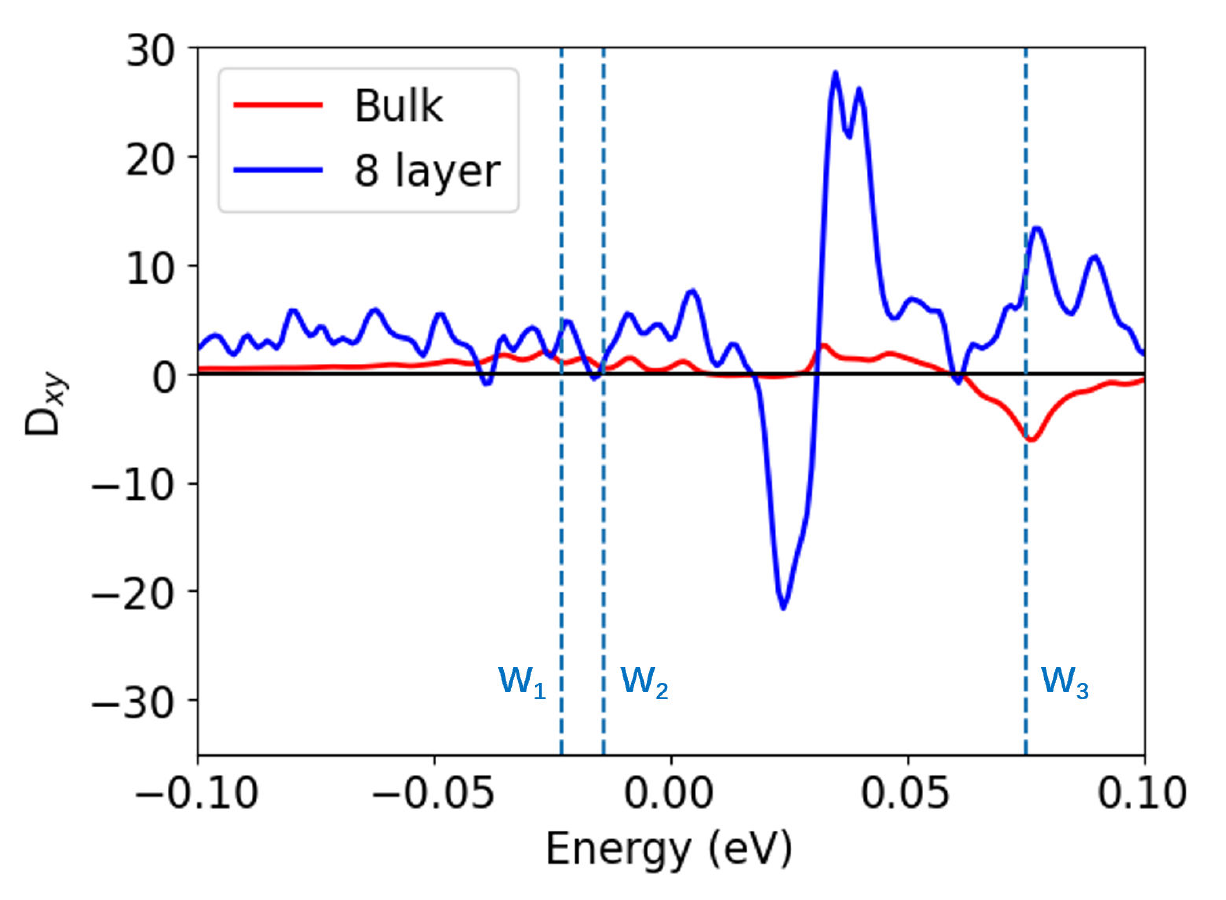}
		\caption{Comparison of BCD $D_{xy}$ in bulk and 8-layer TaAs. The vertical lines indicate the energy levels of W$_1$, W$_2$, and W$_3$ Weyl points in bulk TaAs.}
		\label{fig:bulk-slab}
	\end{figure}
	
	We  calculate BCD of bulk TaAs, and the results are shown in Fig.~\ref{fig:bulk-slab},
	which are consistent with those of previous work \cite{THz}, but slightly different from those of Ref.~\cite{TaAs-BCD}.
	The BCD in Ref.~\cite{TaAs-BCD} has a negative peak around W$_1$, which is absent in this Ref.~\cite{THz} and in this work.
	This is probably due to that the BCD in this energy range
is very small, and sensitive to the calculation details, including the $k$-points meshes and smearing parameters etc.
	The BCD near the Fermi surface is relatively small, primarily attributable to the contributions from the W$_1$ and W$_2$ points~\cite{TaAs-BCD}.
	This finding is intriguing given the significant Berry curvature magnitudes around the Weyl points.
	The reason for the small BCD has been analyzed in Ref. \cite{TaAs-BCD}.  The W$_1$ and W$_2$ points in TaAs are type-I Weyl points, characterized by slightly tilted Weyl cones. As a consequence, a pronounced cancellation effect occurs between the Berry curvatures associated with these Weyl points. To quantitatively assess this cancellation, we introduce a mutual cancellation ratio, which is calculated by taking the absolute value of the BCD and dividing it by the integral of the absolute value of BCD density  at each $k$ point across the entire Brillouin zone (BZ)  (see Eq. \ref{eq:chi} in Sec.\ref{sec:methods} ). A small $\chi$ value indicates significant canceling effects. Remarkably, the ratios around type-I Weyl points W$_1$ and W$_2$ are found to be 0.25 and 0.23, respectively.
In contrast, the cancellation ratio is $\chi$=0.42 for the type-II Weyl points W$_3$.
	
Our analyses further  reveals that,   apart from the inherent cancelation effects within individual Weyl points,
substantial canceling effects emerge from distinct Weyl point pairs in bulk TaAs.
More specifically,  while contributions from the Weyl points  connected by the $M_x$ and $M_y$ symmetries
serve to amplify the BCD, the BCD arising from different groups of Weyl points demonstrates a pronounced tendency for cancellation.
Notably, at 25 meV below the Fermi level, the BCD at W$_1$ has a negative value, whereas the $k$ points around W$_2$ and W$_3$ Weyl points exhibit positive BCD. Consequently, a significant cancellation effect arises among different groups of Weyl points,
resulting in an extremely small cancellation ratio $\chi$ = 0.06 at this energy level, therefore leading to a small BCD.
The same scenario occurs for W$_2$, where a small BCD is also due to cancellation effects. On the other hand,  the type-II W$_3$ Weyl points, are far above Fermi level, and the cancellation effects from W$_1$ and W$_2$ are relatively smaller, resulting in a much larger BCD value,  approximately reaching 7. Nevertheless, it is important to note that W$_3$ resides at a substantially higher energy  than the Fermi level,  posing a challenge for experimental  observation.
		
	\subsection{Berry Curvature Dipole  in TaAs Slabs}
	
	As elucidated in the preceding discussion, the diminished BCD near the Fermi level in bulk TaAs arises primarily from the substantial cancellation effects among different groups of Weyl points. In light of this understanding, our objective is to enhance the BCD in the vicinity of the Fermi energy. A viable approach to achieving this enhancement entails mitigating the canceling effects attributed to the W$_2$ and W$_3$ points. This can be accomplished by selectively excluding the W$_2$ and W$_3$ Weyl points from the energy window in the vicinity of the Fermi level, thereby retaining only the W$_1$ points.
	
	Thin films offer a promising approach for manipulating energy bands through quantum confinement effects. To describe the energy bands in thin films, we may employ the truncated crystal approximation (TCA) \cite{truncated1, truncated2}, which allows us to approximate the energy bands of the thin film, grown along the $z$ axis, at $\boldsymbol{k}=(k_x, k_y)$ using the energy bands of the bulk material at $(k_x, k_y, k_z)$, where $k_z$ takes the values $k_z = n/N_L$ (in units of $2\pi/c$) for $n=1, 2, \dots$ $N_L$, and $N_L$ represents the number of unit cells along the $z$ axis in the thin film. Since the values of $k_z$ for W$_1$, W$_2$, and W$_3$ are 0.00, 0.408, 0.452 respectively,
	by carefully choosing the number of unit cells in the thin film ($N_L$), we can ensure that the bands at $k_z = 0$ are preserved, while the energy bands corresponding to other $k_z$ values will not appear in the thin film bands. This controlled tuning of energy bands therefore has the potential to enhance the BCD around the Fermi energy and enable the realization of desired electronic properties in thin films.
	
	
	\begin{figure}[tbp]
		\centering
		\includegraphics[width=0.5\textwidth]{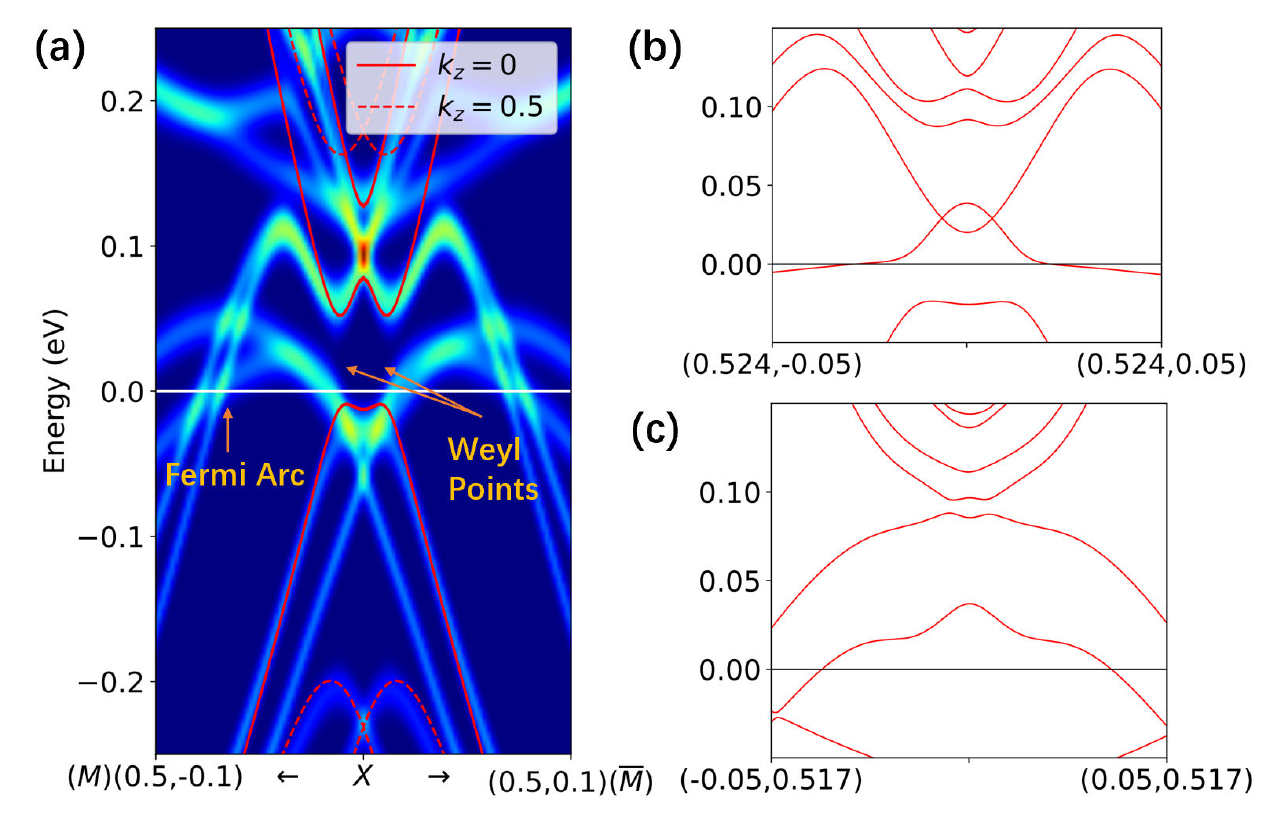}
		\caption{(a) The band structures of the 8-layer TaAs slab from first-principles calculations compared with the band structures of TCA along $M-X-\bar{M}$ near $X$ points. (b) Weyl points near the $X$ point remain gapless.
(c) The Weyl points near the $Y$ point are gapped in the slab.}
		\label{fig:slab-band}
	\end{figure}
	
	To validate this concept, we conduct electronic structure calculations and obtain the band structure and BCD for thin films consisting of 1 to 10 layers of TaAs. Specifically, Fig. \ref{fig:slab-band}(a) illustrates the energy bands of an 8-layer slab (corresponding to $N_L$=2 stacking unit cells along the $z$ axis) around the $X$ point in the two-dimensional BZ. For comparison, we also plot the energy bands of bulk TaAs at $k_z = 0$ and $k_z = 0.5$, which correspond to the band structures of the 8-layer TaAs from TCA.  The band structures of the slab are in relatively good agreement with the bands obtained from the TCA, especially the bands from $k_z$=0 near the Fermi level.
 The energy bands of $k_z$=0.5 are far away from the Fermi level.
	
Remarkably, The W$_1$ Weyl point, located at ${\boldsymbol{k}} = ( 0.517, 0.007, 0)$ near $X$ point in bulk, is still presented in the slab energy bands, as seen in Fig.~\ref{fig:slab-band}(b), which is however at 30 meV above the Fermi level, due to the quantum confinement effects. On the other hand W$_1$ Weyl point near $Y$ point open a small gap, due to the symmetry breaking from C$_{4v}$ to C$_{2v}$.
There are additional surface states, namely Fermi arcs, connecting the W$_1$ points with their opposite chirality (see Fig S2.
in the Supplement materials (SM) \cite{SM} for details).
The Fermi surface structure is consistent with a previous study of 28 TaAs layers\cite{TaAs-fermiarc} for both Ta and As terminations.

Remarkably, the W$_2$ and W$_3$ Weyl points are completely eliminated because their $k_z$ are far from 0 and 0.5.
It is worth noticing that previous studies have reported the existence of W$_2$ Weyl points in TaAs thin films~\cite{TaAs-experiment,TaAs-surfacestate}.
However, these experimental investigations focused on considerably thicker films compared to the ones used in this work.
In those thicker films, the ratio $n/N_L$ can approach the value of $k_z$ for the W$_2$ points, when $N_L$ is large.

\begin{figure}[tb]
		\includegraphics[width=0.45\textwidth]{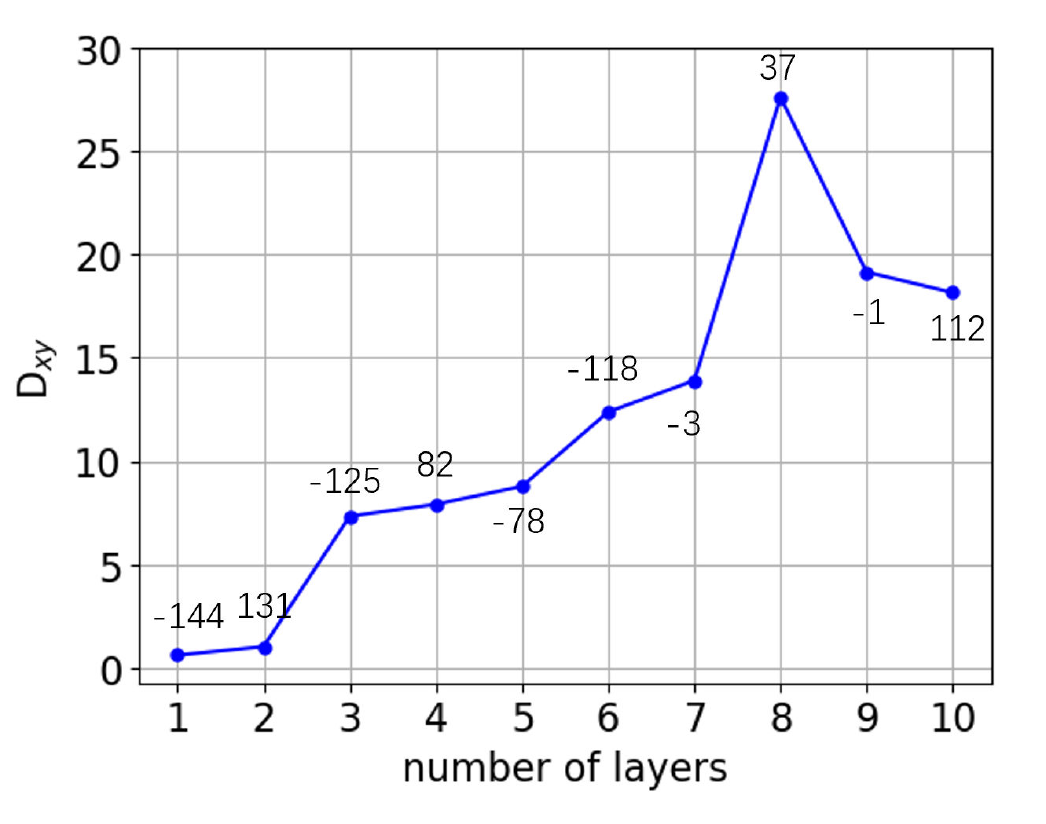}
		\caption{The maximal normalized BCD ($D_{xy}$) of each slab contains different TaAs layers within $\pm$200 meV of the Fermi level.
The number at each point indicates the energy position relative to the Fermi energy in meV.}
		\label{fig:slab-all}
	\end{figure}

Due to the absence of W$_2$ and W$_3$ in the band structure, we anticipate a significantly larger BCD near the Fermi level compared to bulk TaAs, because there are no cancelling effects from these Weyl points.
To further verify our theory and designing principle,  we calculate the BCD near the Fermi energy of 1-10 layers of TaAs slabs, and the results are shown in Fig. S3 - S6 in SM \cite{SM}.
The BCDs are normalized by the thickness of the slab,  $D_{xy} = \frac{2\pi}{L}D^{\rm slab}_{xy}$ where $L$ represents the thickness of the slab.
This guarantees that $D_{xy}$ converges to its bulk value, as $L$ approaches infinite.
Figure \ref{fig:slab-all} shows the maximal  values of  $D_{xy}$ of TaAs slabs within the energy window of $\pm$200 meV around the Fermi level, and the numbers in the figure show the energy with the maximal BCD related to the Fermi level. As expected, the 8-layer TaAs slab has largest layer normalized BCD, which are significantly larger than the BCD of other slabs.
	
Figure \ref{fig:bulk-slab} compares the BCD of the 8-layer TaAs slab to that of bulk TaAs material within $\pm$ 100 meV around the Fermi level. Notably, the BCD values of the 8-layer TaAs slab are significantly higher than those of the bulk counterpart.
	The 8-layer TaAs slab has two peaks at approximately 25 meV and 37 meV above the Fermi level, and which have BCD values
	of -20.0 and 27.6, respectively.
	These values are not only much larger than that of bulk TaAs at 75 meV contributed from W$_3$, more importantly, the
	corresponding energies are also much closer to the Fermi level.
	These energies are reasonably reachable through traditional Fermi surface tuning methods, such as, doping or applying external electric voltage. These energy levels are within reasonable reach through conventional Fermi surface tuning methods such as doping or the application of external electric voltage.
		
	\begin{figure}
		\centering
		\includegraphics[width=0.35\textwidth, ]{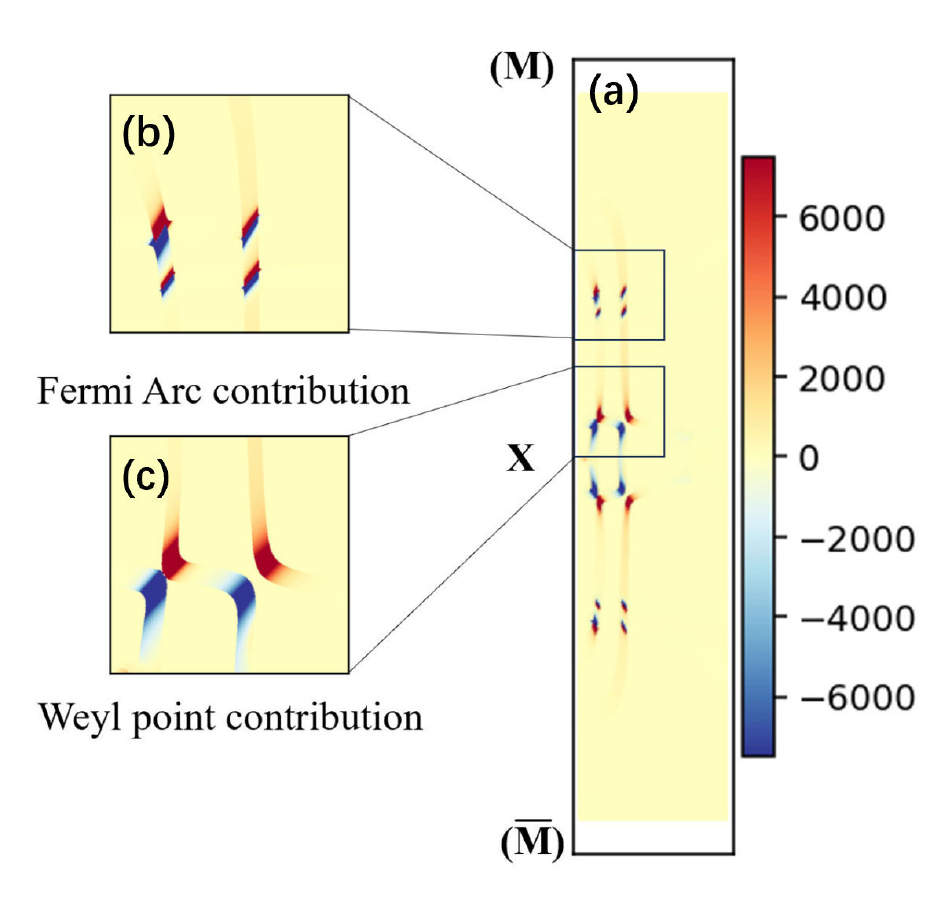}
		\caption{(a)The BCD distribution in the 8-layer TaAs near the $X$ point. (b) The BCD distribution from the Fermi arc.
(c) The BCD distribution from the  W$_1$ Weyl point.  }
		\label{fig:slab-distribution}
	\end{figure}

To further explore the origin of large BCD in the 8-layer slab, we analyze contribution to the BCD in the BZ. We compare the BCD distribution in bulk and in 8-layer slab TaAs on the $k_xk_y$ plane, focus on the energy  at W$_1$ Weyl point,  i.e., 23 meV below Fermi level for bulk and 37 meV above Fermi level for the slab. Both distribution patterns are sampled in a 10 meV energy window around the targeted energy. More details can be found in Fig. S3 of \cite{SM}.
The bulk results are in good agreement with previous work\cite{TaAs-BCD}.
In bulk TaAs, there are substantial contributions from W$_1$, W$_2$, and W$_3$, with opposite signs.
However, in the slab system, contributions from W$_2$ and W$_3$ vanish, leaving only the contribution around the W$_1$ points.
This highlights  the effectiveness of our design strategy employing slabs to resolve the issue of canceling effects among distinct groups of Weyl points.

Figure \ref{fig:slab-distribution} depicts the significant contributions around the W$_1$ point near X.
The $k$-points near Weyl points and Fermi arcs exhibit a very high BCD density, reaching magnitudes of 10$^4$, whereas contributions from other $k$-points are negligibly small.
Furthermore, the BCD density in the slab is significantly larger than that of the bulk system, primarily due to the substantial enhancement of the density of states resulting from quantum confinement effects.

The slab hardly change the tilting of Weyl cone, thus does not improve the canceling rate within individual W$_1$ points. In fact, $k$-points near slab W$_1$ has a canceling rate of $\chi$=0.20, slightly worse than that of the bulk.

In addition to the contribution from W$_1$, the Fermi arc also makes a significant contribution to the BCD, as illustrated in Fig.~\ref{fig:slab-distribution}.
The Fermi arc accounts for 21.5 out of the total BCD of 27.6.
As the Fermi arcs all originate from W$_1$ points, there is no cancellation effect from Fermi arcs originated from W$_2$ and W$_3$ points.
The cancellation ratio for the Fermi arc is 0.23, which leads to an overall $\chi$ value of 0.224 for the 8-layer slab,
 in contrast to the 0.06 value observed in bulk TaAs.

As shown in Fig.~\ref{fig:slab-all}, the maximal value of $D_{xy}$ of the slab increases with the number of layers and reach maximal values at 8-layer slab, and the value then decreases. This is consistent with the analysis above. In the case of a 4-layer slab, corresponding to $N_L$=1, the band structure should also include W$_1$ points, as predicted by the TCA. However, for this very thin slab,  the coupling between the upper and lower surfaces becomes more stronger, the TCA may not be a good approximation.
We also compute the BCD for both 12-layer and 16-layer TaAs slabs. The maximum BCD values are 18.38 and 16.05, respectively.

\section{Summary}
	
We have developed an experimentally feasible strategy for enhancing the BCD in TaAs materials using thin films, capitalizing on the benefits offered by quantum confinement effects.
In addition to enhancing the density of states, a critical concept lies in the precise selection of the thin film's thickness, to retain desired $k$-points at the Fermi surface while simultaneously excluding unwanted ones through quantum confinement effects,  and therefore minimizes the canceling effects among these $k$-points.
This strategy extends beyond the current research for TaAs BCD enhancement, offering a versatile approach to selectively choose desired $k$-points for diverse applications in various materials and systems.

\acknowledgements
	This work was funded by the Chinese National Science Foundation Grant Numbers 12134012. The numerical calculations were performed on the USTC HPC facilities.

%

\end{document}